\documentclass{elsarticle}
\usepackage[utf8]{inputenc}
\usepackage{amsmath}
\usepackage{graphicx}
\usepackage{color}
\usepackage{abbrev}
\usepackage{amssymb}
\begin{document}
\begin{frontmatter}
\title{The linear viscoelastic fracture theory applies to soft solids better when they are... viscoelastic.}
%\title{Crack growth in viscoelastic solids -- can linear models stop failing to account for the experimental results?}
%or//
%Another theory of viscoelastic crack in the frequency domain

\author{Etienne Barthel}
\address{Soft Matter Sciences and Engineering, ESPCI Paris, PSL University,
Sorbonne Universit\'{e}, CNRS, F-75005 Paris, France}
\date{\today}
%
%Central idea : examples for fracture - give one for adhesion
%
\begin{abstract}
Over the last half-century, linear viscoelastic models for crack growth in soft solids have flourished but their predictions have rarely been compared to experiments. In fact, most available models are either very approximate or cast in forms which are not quite suitable for the analysis of actual data. Here, we propose a linear viscoelastic approach which consistently exploits the dynamic mechanical analysis (DMA) data. We apply this method to four sets of results documenting fracture or adhesion rupture in soft solids with various degrees of viscoelasticity. For elastomers, the results reproduce the well-known inconsistency of the process zone size. In more viscoelastic systems however, the present approach is able to match the measured velocity dependence of the rupture energy with physically acceptable process zone sizes. Moreover, our predictions agree with the damage zone sizes measured by mecanoluminescence. Building on these results, we discuss various issues arising when evaluating the linear viscoelastic contribution to the rupture/adhesion energy in soft solids: data quality, physical interpretation of the parameters, validity of simpler approximations and limitations of the present approach.
\end{abstract}

\end{frontmatter}

\section{Crack growth in viscoelastic solids - the linear model}

In the rupture of soft polymers (elastomers, gels, viscoelastic polymers) the effective rupture energy $G$ compounds the interfacial rupture energy $w$ properly speaking and the energy expended through bulk material dissipation around the crack tip. Remarkably, it has been found that the rupture energy scales with velocity $V$ and temperature using the time-temperature scaling~\cite{Gent94,Gent96} derived from the linear dynamic mechanical response. Although fracture or adhesion rupture inevitably involve large strains near the crack tip, especially in soft materials, this result suggests that linear viscoelasticity dominates the dissipation processes during crack growth. In a scheme where the non linear processes (viscoelastic non linearities, damage, rupture,...) are assumed to be confined to a process zone around the crack tip, the energy dissipated by viscoelasticity in a region surrounding this process zone can be calculated by a linear theory and the results can be used to predict the evolution of the effective fracture/adhesion energy as a function of velocity. And indeed many such theories have been developed over the years (see \cite{Knauss15_reviewfractureviscoelastic,Hui22} for reviews).

In parallel, in the field of soft matter, fracture data have often been analysed as follows. Power law exponents for a) the fracture/adhesion energy increase with crack velocity and b) the loss function increase with frequency are compared (see~\cite{Sun17_Bulkenergydissipation} for an example). Because of the rather loose idea that the enhancement of fracture energy with velocity should be related to the increase of the loss modulus with frequency, similarity of these two exponents is deemed sufficient to demonstrate that linear viscoelastic theory applies and, to that point, one could think that the fracture problem in soft materials is settled. In fact, despite the abundance of theories on the one hand and data on the other hand, we find very few in depth comparisons of the ones with the others. The first problem is that, of the many theories available, very few are cast under a form suitable for comparison with experiments. The second problem is that the data are often incomplete and not amenable to an in depth analysis. The third problem is that, when a full comparison is finally made, there always appears a serious inconsistency. One of the prime examples is Gent's study of rupture/adhesion energies for several types of elastomers and interfaces. Using a simple linear viscoelastic approach, he evidenced a characteristic size for the non linear process zone of the order of 0.1~nm~\cite{Gent94,Gent96}. More elaborate forms of the linear viscoelastic theory by Hui~\emph{et al.}~\cite{Hui22} come to identical conclusions~\footnote{Ref. \cite{Hui22} also brings into focus the paucity of good data -- the two sets of data they analyzed were respectively a quarter and half a century old... }. These values for the process zone size are typical for viscoelastic fracture theory applied to elastomers : they are unacceptably low and this shortcoming critically challenges the applicability of the theory.

In the present work, we attempt to bridge the gap between experiments and theory. We propose a simple expression relating the dynamic storage modulus measured as a function of frequency to the velocity dependence of the rupture/adhesion energy. We then apply this expression to four different sets of data and show that: a) the elastomeric systems do exhibit an implausibly small process zone size; b) more viscoelastic systems have reasonable process zone sizes, in the micron to 10 micron range, which compare favorably with the damage zone sizes measured by mecanolumicescence. In the light of these analyses, we discuss various issues relevant for the evaluation of the contribution of linear viscoelasticity to rupture/adhesion energies and the comparison of these predictions with data.

\section{An interesting experiment}\label{sec:Tay_data}
In 2006, Tay \& Barquins~\cite{Tay06} carried out JKR type adhesion experiments~\cite{JKR} with glass balls pressed against flat surfaces made of natural rubber (elastic modulus in extension E=0.89 MPa, Poisson ratio 0.498, Tg=-72$^{\circ}$~C). They determined the effective adhesion energy as a function of velocity $G(V)$. In JKR set-ups, the effective adhesion can be measured over extremely long periods of time: as a result, the limit $w=G(V\simeq 0)$ could be measured precisely, which is a key point for data analysis (see Sec.~\ref{sec:discussion}). Quantitatively, they found a dependence $\phi=(G(V)-w)/w=k(a_T V)^{0.55}$ where $w=42.7$~mJ/m$^2$, $k\simeq 45$, $a_T=1.570\times10^{-3}$ at 25$^\circ$C and the velocity is given in $\mu$ms$^{-1}$. With V in ms$^{-1}$, this is
\begin{equation}\label{eq:phi_msm1}
\phi(V)=s\ V^{0.55}    
\end{equation} with $s=2.57\times10^3$ in SI units. This power law, which applies over three decades of velocity, clearly reflects the role of the dissipative processes in the glass/elastomer rupture.

\begin{figure}
\begin{center}
\includegraphics[width=10cm]{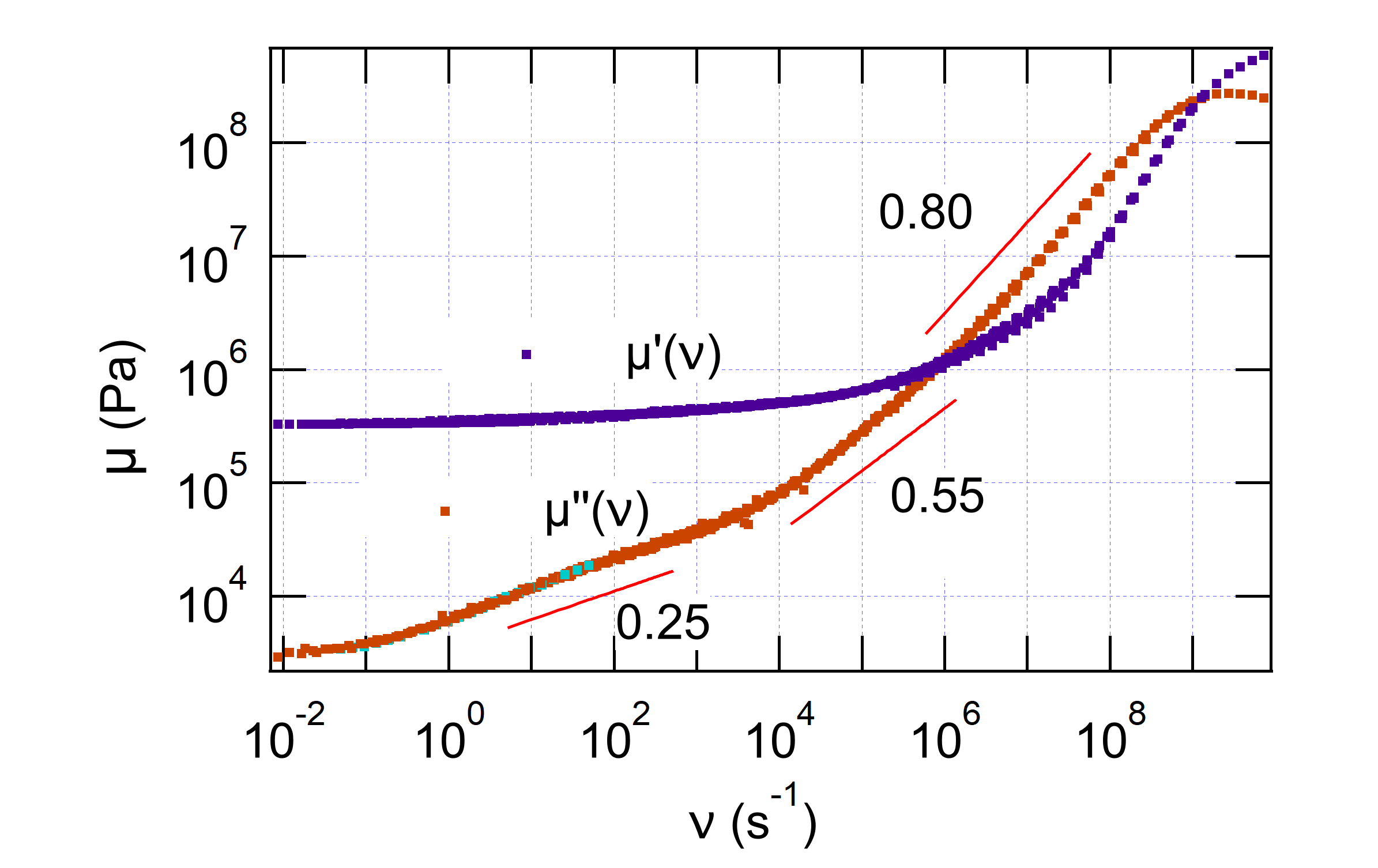}
\caption{Storage modulus $\mu'(\nu)$ and loss modulus $\mu''(\nu)$ for the natural rubber used in~\cite{Tay06}, referred to room temperature. Local power law dependencies over frequency are shown for the loss modulus.}\label{fig:Tay_GpGpp}
\end{center}
\end{figure}

We want to compare this result with predictions from a linear viscoelastic theory. For polymers the dynamic, small strain, linear response is usually measured by dynamic mechanical analysis (DMA) as a function of frequency $\nu$. For the natural rubber used by Tay \& Barquins, the storage modulus $\mu'(\nu)$ and loss modulus $\mu''(\nu)$ are shown in Fig.~\ref{fig:Tay_GpGpp}, referenced to room temperature.

% -------------------------------------------------------
% ------------------------------------------------------

\section{The model}
We now show that a useful expression to calculate $G(V)$ from DMA data can be derived from our earlier solution to the linear viscoelastic crack problem~\cite{Barthel09}.

\subsection{Solution in the time domain - effective crack tip compliance}
\begin{figure}
\begin{center}
\includegraphics[width=8cm]{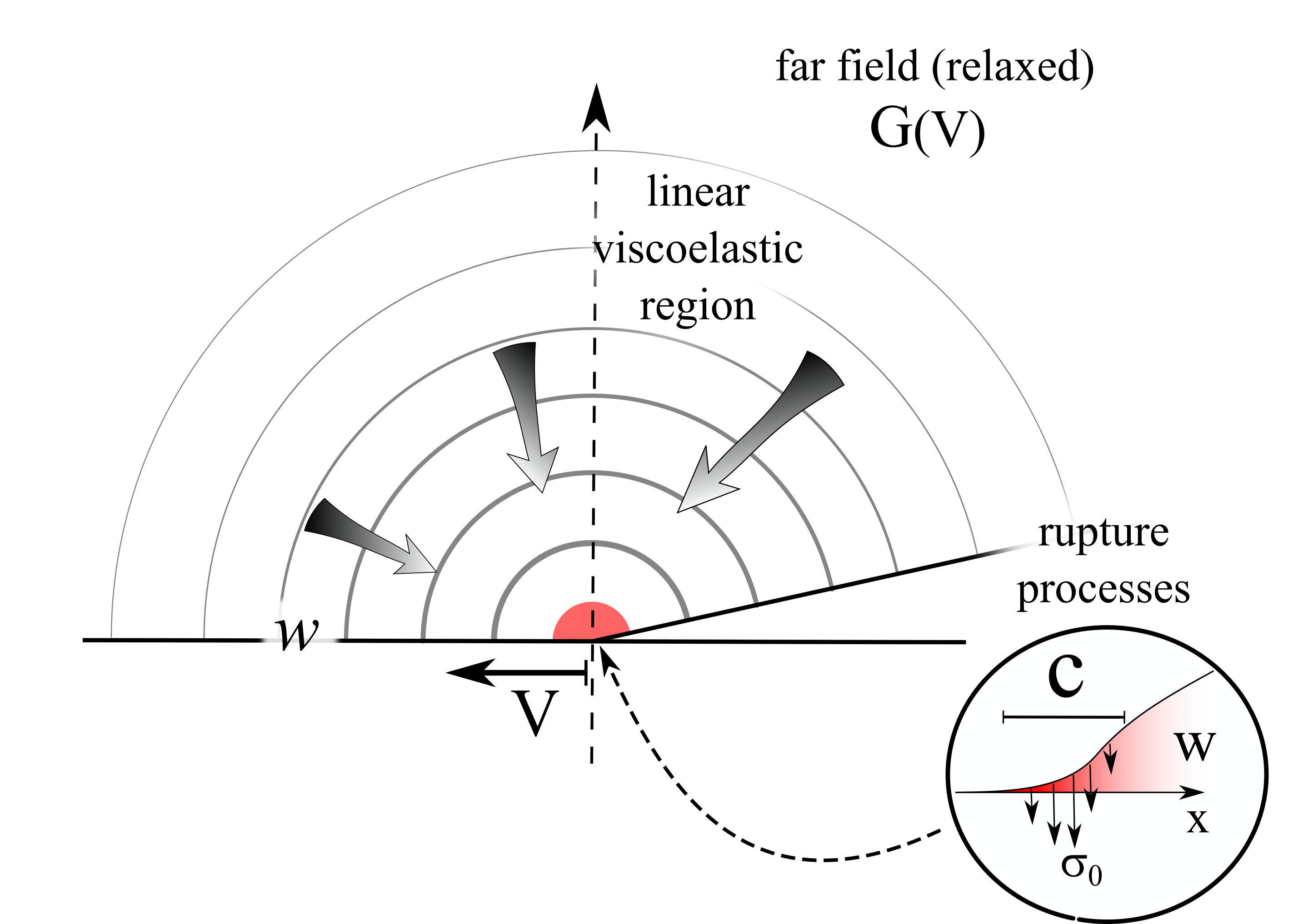}
\caption{Schematics of the energy flux from the relaxed far field through the dissipative viscoelastic region to the crack tip (in red). The crack is moving at velocity $V$. To model the contribution of the viscoelastic dissipation to the effective, far field, rupture/adhesion energy $G(V)$, the rupture processes themselves are modelled by a cohesive zone (inset, in red). The interfacial rupture energy for face separation is $w$ and the cohesive stress $\sigma_0$. The size of the cohesive zone $c(V)$ depends upon $w$ and $\sigma_0$ and the material viscoelastic response through Eq.~\ref{eq:cohesive_zone_viscoelastic}.}\label{fig:cohesive_zone_schematics}
\end{center}
\end{figure}

In a nutshell, we assume that the region over which the material separates, the process zone, extends over a characteristic size $c$ and that this separation induces tensile stresses of magnitude $\sigma_0$ (the cohesive stress) over this region. With this so-called cohesive zone model we can couple the rupture processes at the crack tip to the linear viscoelastic dissipation in the surrounding material (Fig.~\ref{fig:cohesive_zone_schematics}). It is similar to methods previously propounded by Greenwood~\cite{Greenwood04} and many others~\cite{Knauss15_reviewfractureviscoelastic,Hui22}.  
% ---------------------------------------------------------
% -----------------------------------------------------------

Starting with a purely elastic material, we use the linear elastic solution to calculate the work of the cohesive stress in the process zone~\cite{Haiat07}. Since this work is equal to the adhesion energy $w$, we obtain
\begin{equation}\label{eq:Cohesive_Zone_Size_Elastic_}
w=\frac{\pi}{4}\frac{{\sigma_0}^2c}{{E}^\star}\
\ \ \ \ \hbox{(linear elastic)}
\end{equation}
where $\Es$ is the plain strain modulus. The approximate derivation sketched in~\ref{sec:appendix_self_con} shows that Equation~\ref{eq:Cohesive_Zone_Size_Elastic_} is a {\em self-consistency relation} for the stress and strain distributions around the process zone.

Turning now to viscoelastic materials, we assume a steady state propagation at velocity $V$. The process zone size $c$ now depends upon $V$ and we introduce a characteristic time $t_r$~\cite{Greenwood81} such that
\be\label{eq:definition_V}
c(t_r) = V t_r
\ee
This time will be called here {\em transit time}, although the phrases residence time or dwell time have also been used. The transit time sets the characteristic strain rate around the crack tip.

The distribution of strain rates in the region surrounding the process zone can be taken into account through the standard method of linear viscoelasticity, \emph{i.e.} time convolution of elastic solutions. We can calculate the work of the cohesive stress for a steady state propagation of the process zone~\cite{Haiat07}. In this time domain description, using the viscoelastic compliance $J(t)$ of the material, we find that the work of the cohesive stress is proportional to an effective crack tip compliance $J_1(t_r)$. In a moving frame picture, $J_1(t_r)$ subsumes the response of the viscoelastic material flowing around the process zone of size $c$ with velocity $V$ (Fig.~\ref{fig:cohesive_zone_schematics}). The expression for $J_1$ derived in \cite{Barthel09} is given in~\ref{sec:appendix_frequency_domain} (Eq.~\ref{eq:Effective_Compliance_Opening}). Equating again the work of the cohesive stress to the interfacial rupture energy $w$, we obtain the self-consistency equation~\cite{Barthel09}
\begin{equation}\label{eq:cohesive_zone_viscoelastic}
w=\frac{1}{2}{\sigma_0}^2 c(t_r) {J_1}(t_r)\ \ \ \ \ \hbox{(linear viscoelastic)}
\end{equation}
Equation~\ref{eq:cohesive_zone_viscoelastic} generalizes Eq.~\ref{eq:Cohesive_Zone_Size_Elastic_} in that $J_1(t_r)$ replaces $1/E^*$.

However, this equation on its own is not sufficient to predict the effective rupture/adhesion energy $G(V)$, which is measured in the far field. If we assume the system is large enough, \emph{i.e.} there is no finite size effects, then far from the dissipative region surrounding the process zone, the material is in a fully relaxed state (Fig.~\ref{fig:cohesive_zone_schematics}) and the far field response is described by $J(t=\infty)$, the long time compliance. As a result~\cite{Barthel09}, the effective rupture/adhesion energy can be expressed as
\begin{equation}\label{eq:effective_adhesion_receding}
\frac{G(V)}{w} = \frac{J_1(t=\infty)}{J_{1}(t_r)} 
\end{equation}
where we have used $J(t=\infty)=J_1(t=\infty)$.
% -------------------------------------------------------------
\begin{figure}
\begin{center}
\includegraphics[width=10cm]{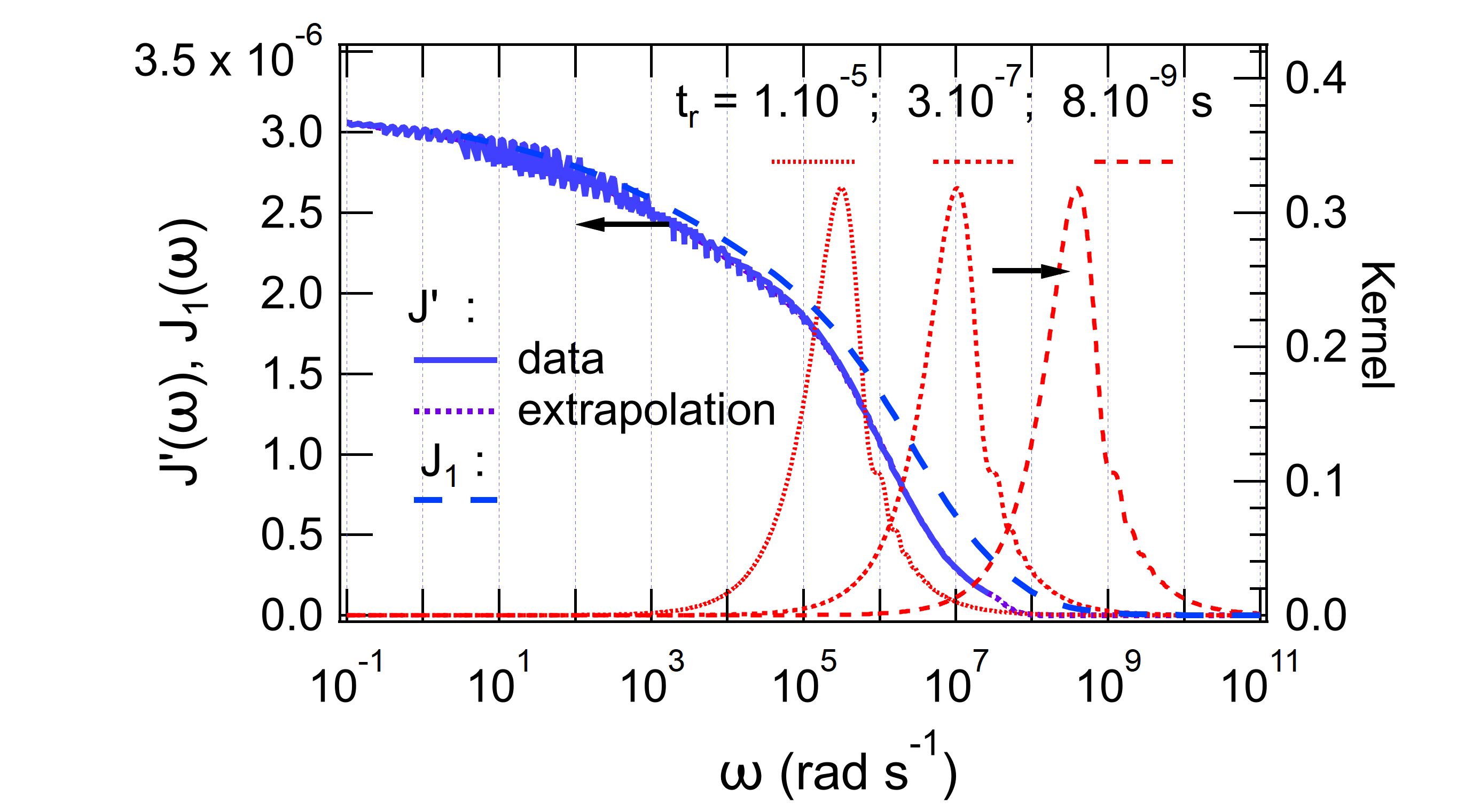}
\caption{Calculation of the effective crack tip compliance $J_1(t_r)$ for transit time $t_r$ using Eq.~\ref{eq:J1_freq} and the data given in Fig.~\ref{fig:Tay_GpGpp}). The storage compliance $J'(\omega)$ is shown (extrapolated above $\omega=2.5\ 10^7$ rad.s$^{-1}$) along with the kernel for three different values of $t_r$. The calculated $J_1$ is also plotted as a function of $\omega=2\pi/t_r$, for comparison with $J'$.}\label{fig:Jp_J1}
\end{center}
\end{figure}
% -----------------------------------------------------------
\subsection{Effective crack tip compliance in the frequency domain}
Eqs.~\ref{eq:cohesive_zone_viscoelastic} and \ref{eq:effective_adhesion_receding} provide the required relations between the rupture/adhesion energy enhancement $G(V)/w$ and the linear viscoelastic compliance $J(t)$. This time domain solution, however, is inconvenient because the dynamic response is usually measured by DMA, in the frequency domain. Also, compared to time domain measurements, the DMA data usually span a wide range of frequencies through time-temperature superposition, including the very high frequencies necessary for the analysis of elastomeric materials, as seen in Fig.~\ref{fig:Tay_GpGpp}. To obtain an expression for the effective crack tip compliance $J_1$ in the frequency domain, we use the complex compliance $J(\omega)$, which is the Fourier transform of $J(t)$. As shown in \ref{sec:appendix_frequency_domain}, $J_1$ can be expressed as a function of the storage compliance $J'$, the real part of  $J(\omega)$:
\begin{equation}\label{eq:J1_freq}
J_{1}(t_r) = \frac{4}{\pi}\int_0^{+\infty} J'(\omega) \frac{\omega t_r - \sin(\omega t_r)}{\omega^2{t_r}^2}d\left(ln(\omega)\right)
\end{equation}
Finally, $J(\omega)$ is the inverse of the complex shear modulus $\mu(\omega) = \mu'(\omega)+i\mu''(\omega)$ which is directly measured by DMA:
\begin{equation}\label{eq:J_mu}J(\omega) = J'(\omega)+i J''(\omega)=  \mu(\omega)^{-1}
\end{equation}
\subsection{Using the model}\label{sec:using_the_model}
In practice, we start from the measured moduli $\mu'(\omega)$, $\mu''(\omega)$ (see Fig.~\ref{fig:Tay_GpGpp} for an example). Algebraic inversion of the complex modulus results in the complex compliance $J(\omega)$ and the real part, the storage compliance $J'$, is shown in Fig.~\ref{fig:Jp_J1}. The calculation of the effective crack tip compliance for a any given $t_r$ is carried out through Eq.~\ref{eq:J1_freq} by convolution of $J'$ with the kernel, in a $ln(\omega)$ scale (natural logarithm). Fig.~\ref{fig:Jp_J1} also displays the kernel for three different transit times. Since $t_r$ only appears in the product $\omega t_r$ in Eq.~\ref{eq:J1_freq}, in a log scale the kernel is shifted by $\log(1/t_r)$ with an invariant shape. Interestingly, in contrast to other approaches~\cite{Gennes96}, there is no need for arbitrary cut-offs to ensure convergence of integral expressions: the natural regularization of the crack singularity by the cohesive zone model is reflected in the shape of the kernel (Fig.~\ref{fig:Jp_J1}), which provides high frequency convergence (Eq.~\ref{eq:J1_freq}). The resulting effective compliance $J_1$ is also plotted in Fig.~\ref{fig:Jp_J1} as a function of $\omega=2\pi/t_r$. Note that due to the convolution process, it shows the same evolution as $J'$ but somewhat shifted to higher frequencies.
% ------------------------------------------------------
\begin{figure}
\begin{center}
a)\includegraphics[width=.85\textwidth]{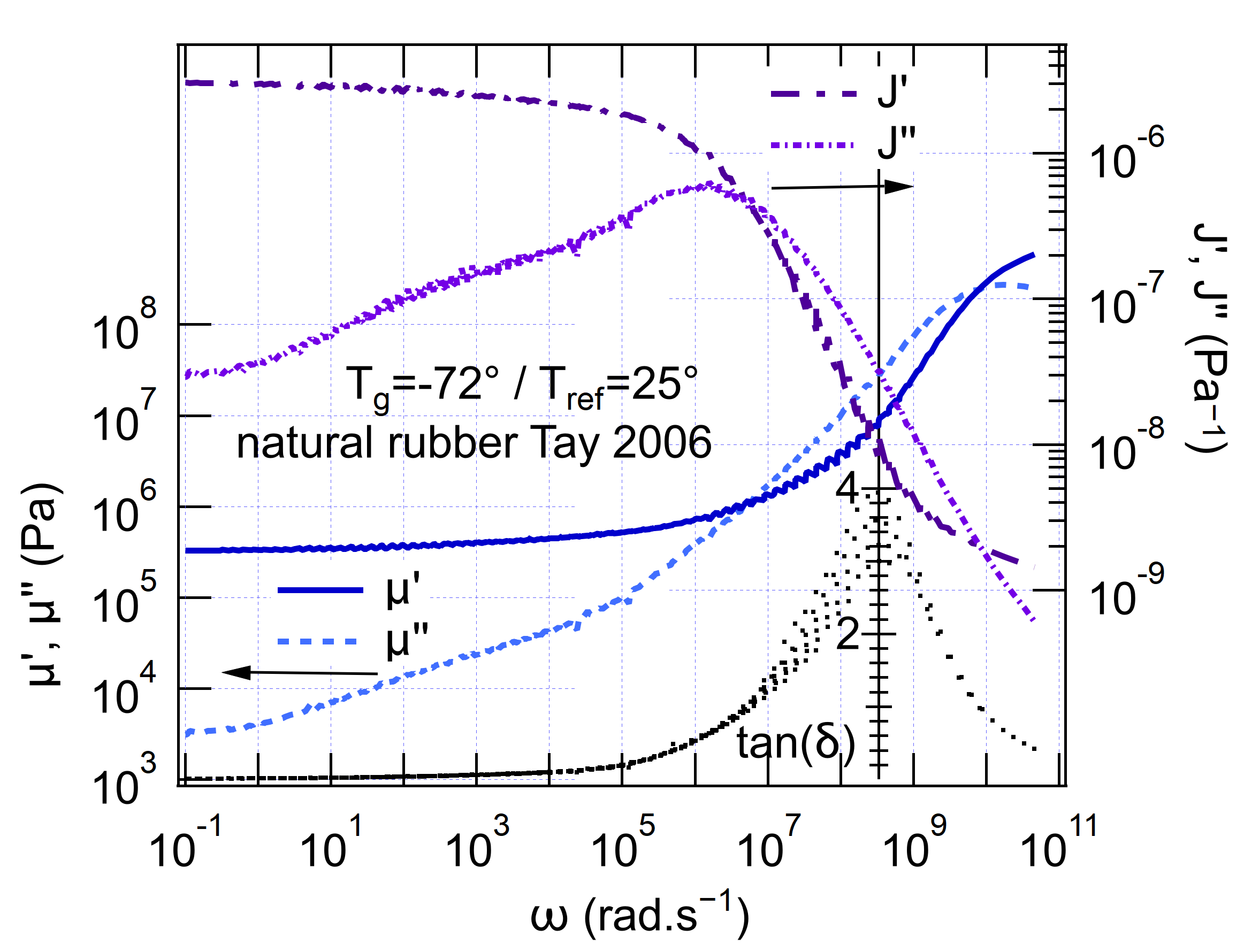}
b)\includegraphics[width=.85\textwidth]{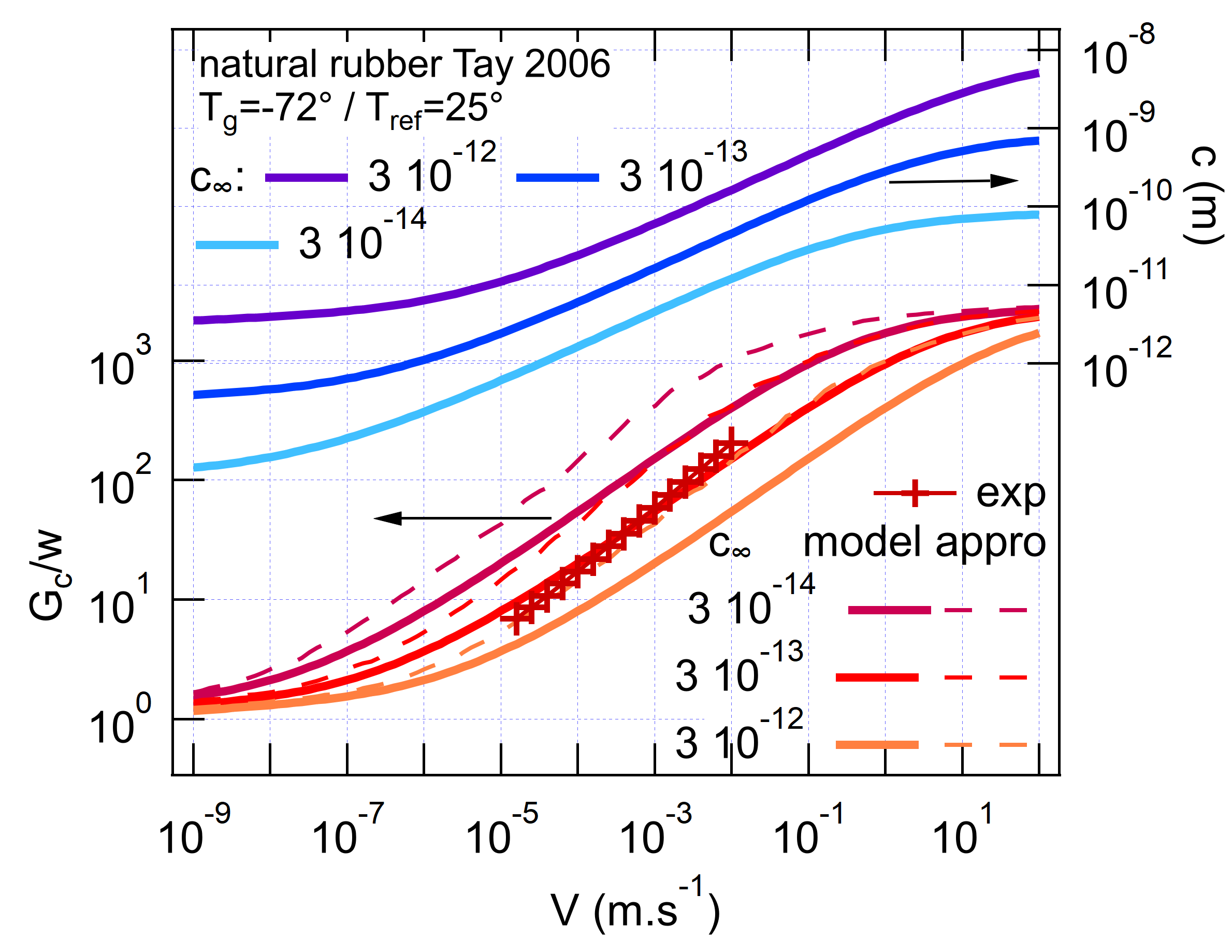}
\caption{Adhesion rupture for natural rubber on glass. a) left axis - real and imaginary parts of the shear modulus as a function of angular frequency  (cf Fig.~\ref{fig:Tay_GpGpp}) - $\tan(\delta)$ is also shown (inner y axis scale). Right axis - real (cf Fig.~\ref{fig:Jp_J1}) and imaginary parts of the shear compliance as a function of angular frequency. b) left axis - adhesion energy as a function of velocity (shown as crosses - calculated from fit to experimental data from~\cite{Tay06}) and models for three values of the single free parameter $c_\infty$. The result for a simpler form of the model (Eq.~\ref{eq:J1_appro}) is also shown as dashed lines. Right axis - evolution of the predicted cohesive zone size as a function of velocity for the full model.}\label{Fig:J_G_Tay}
\end{center}
\end{figure}
% ---------------------------------------------------------

Once the effective compliance has been calculated from the DMA data for all $t_r$, we can simultaneously solve Eqs~\ref{eq:definition_V} and  \ref{eq:cohesive_zone_viscoelastic} for $t_r$ and $c(t_r)$, for any given crack velocity $V$. Finally, we obtain the predicted effective rupture/adhesion energy enhancement $G(V)/w$ from Eq.~\ref{eq:effective_adhesion_receding}.

The limit cases are easily recovered. For smaller velocities $V$ (larger transit times $t_r$) the kernel shifts to lower frequencies, the effective compliance increases, the effective rupture energy $G$ decreases and reaches $w$ for $t_r\rightarrow\infty$. Conversely, for larger velocities $V$ (smaller transit times $t_r$), the kernel shifts to higher frequencies, the effective compliance decreases and the effective rupture energy $G$ increases. For $t_r\rightarrow0$ ($V\rightarrow\infty$), the enhancement reaches \begin{equation}
    J(t=\infty)/J(t=0)=J(\omega=0)/J(\omega=\infty)=\mu'(\omega=\infty)/\mu'(\omega=0)
\end{equation}\emph{i.e.} the ratio of the moduli in the glassy and the elastomeric states.

% -----------------------------------------------------
\section{First examples - elastomers}
% -------------------------------------------------------
As a first example of application, we turn back to the adhesion of natural rubber to glass measured by Tay \& Barquins~\cite{Tay06}. The velocity dependence of $G(V)$ (Eq.~\ref{eq:phi_msm1}) has been measured over three decades and the interfacial rupture energy $w$ has been determined from the very low velocity experiment so that $\sigma_0$ is the only free parameter. Loss and storage modulus have been measured over about 12 decades (Fig.~\ref{fig:Tay_GpGpp}). We have calculated $J'$, $J_1(t_r)$ (Fig.~\ref{fig:Jp_J1} - see also Fig.~\ref{Fig:J_G_Tay} a) and $G(V)$ through Eq.~\ref{eq:effective_adhesion_receding} (Fig.~\ref{Fig:J_G_Tay} b). In practice, we use $c_\infty\equiv c(t_r=\infty)$ as the free parameter, which is directly connected to $\sigma_0$ through Eq.~\ref{eq:cohesive_zone_viscoelastic}. Results for different values of $c_\infty$ are shown in Fig.~\ref{Fig:J_G_Tay} b and we find a reasonable agreement between model and data for $c_\infty=3\ 10^{-13}$~m.  For the evolution of $G(V)$, assuming a power law dependence, the model predicts a slightly lower exponent than observed.  In fact, due to the spread of the kernel, this exponent is still marginally sensitive to the high frequency data for $J(\omega)$ which, above $\omega=2.5\ 10^7$~rad.s$^{-1}$, lies beyond our measurement range (Fig.~\ref{fig:Jp_J1}) and must be extrapolated (\emph{e.g.} by an inverse power function). In any case, the process zone size is excessively small, as typical in the literature.

\begin{figure}
\begin{center}
\includegraphics[width=.85\textwidth]{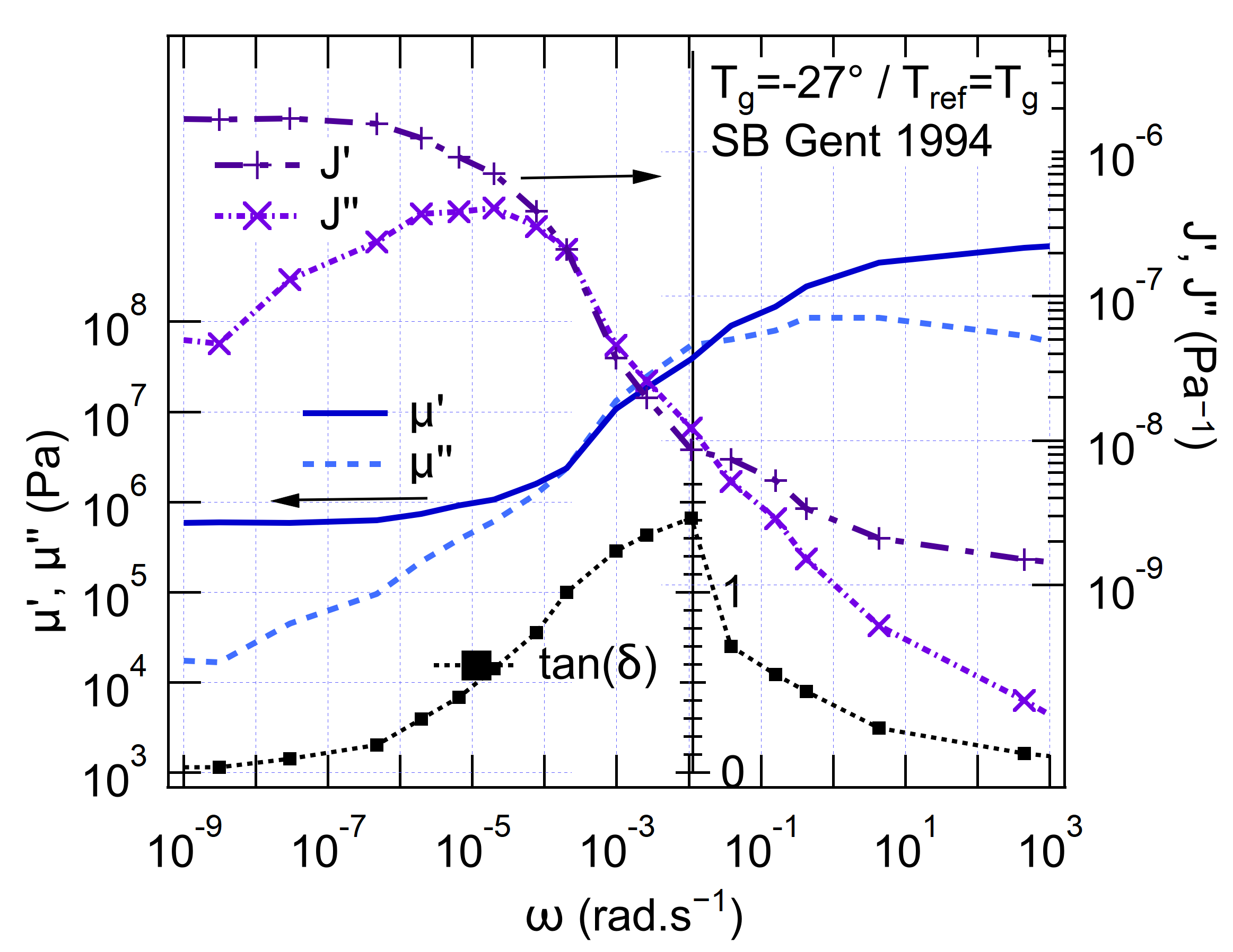}
\includegraphics[width=.85\textwidth]{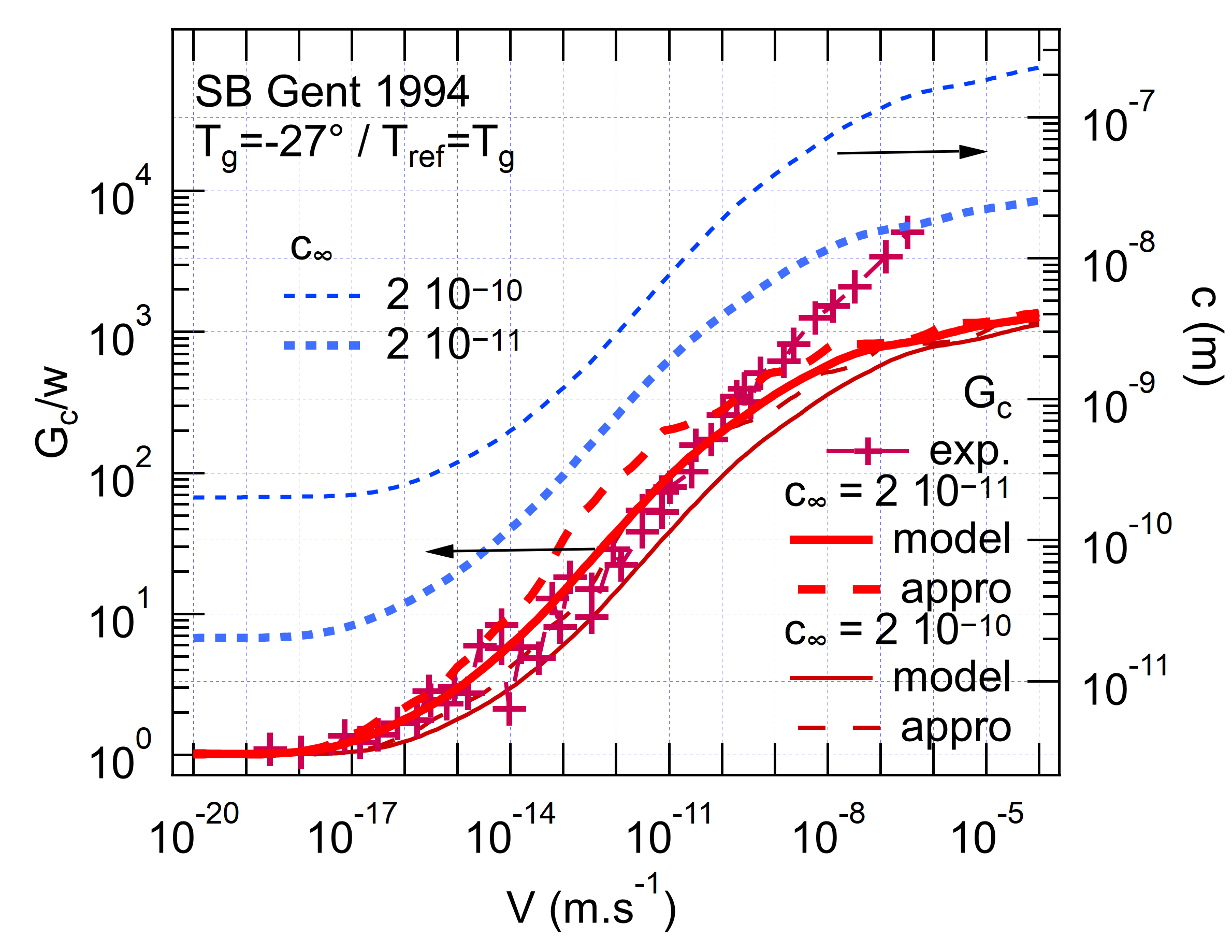}
\caption{Fracture energy of a styrene butadiene rubber. Data taken from~\cite{Gent94}. a) left axis - real and imaginary parts of the shear modulus measured as a function of angular frequency - $\tan(\delta)$ is also shown (inner y axis scale). Right axis - real and imaginary parts of the shear compliance as a function of angular frequency. b) left axis - measured fracture energy as a function of velocity and models for two values of the free parameter $c_\infty$. The result for the approximation Eq.~\ref{eq:J1_appro} is also shown as dashed lines. Right axis - evolution of the cohesive zone size as a function of velocity for the full model.}\label{Fig:J_G_Gent}
\end{center}
\end{figure}

Similarly, we can apply the theory to the elastomer peel data taken from reference~\cite{Gent94}. In this paper, the evolution of $\mu'$ and $\mu''$ with angular frequency is reported (their Fig.~14) along with the evolution of $G(V)$ (their fig.~11 - triangles) for styrene butadiene crosslinked with 1.5~\% peroxide (T$_g$=-27$^\circ$). $G(V)$ exhibits a noticeable plateau at low velocities from which we derive a reasonably reliable value of $w$ = 2.3~J/m$^2$. The data $\mu'(\omega)$, $\mu''(\omega)$ (left) and the compliance $J'(\omega)$ and $J''(\omega)$ (right) are shown in Fig.~\ref{Fig:J_G_Gent} a while Fig.~\ref{Fig:J_G_Gent} b displays the predicted $G(V)$ (left) and $c(V)$ (right). We find a reasonable agreement for $c_\infty=2\ 10^{-11}$~m. Our value is close to the value ($\simeq 1\ 10^{-11}$~m) estimated by the authors (more about their method in the Discussion) and follows a similar evolution with velocity (their Fig.~17). As in the previous example, the process zone size is exceedingly small, which seems to confirm that the linear viscoelastic theory is actually unsuitable to account for the fracture or adhesion energy in elastomers.

% -----------------------------------------------
\section{Analysing data for more viscoelastic systems}
% ------------------------------------------------

\begin{figure}
\begin{center}
a) \includegraphics[width=.85\textwidth]{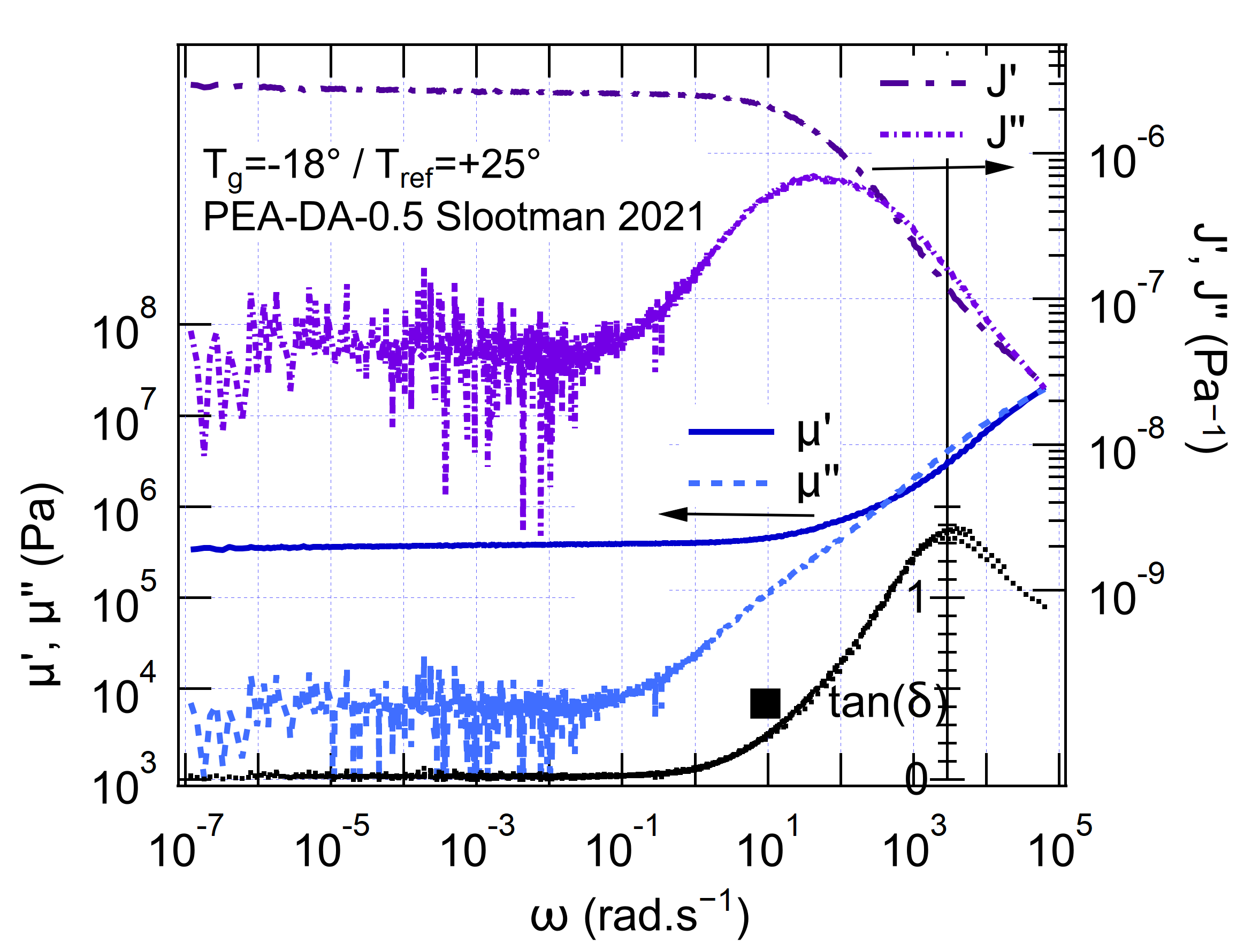}

b) \includegraphics[width=.85\textwidth]{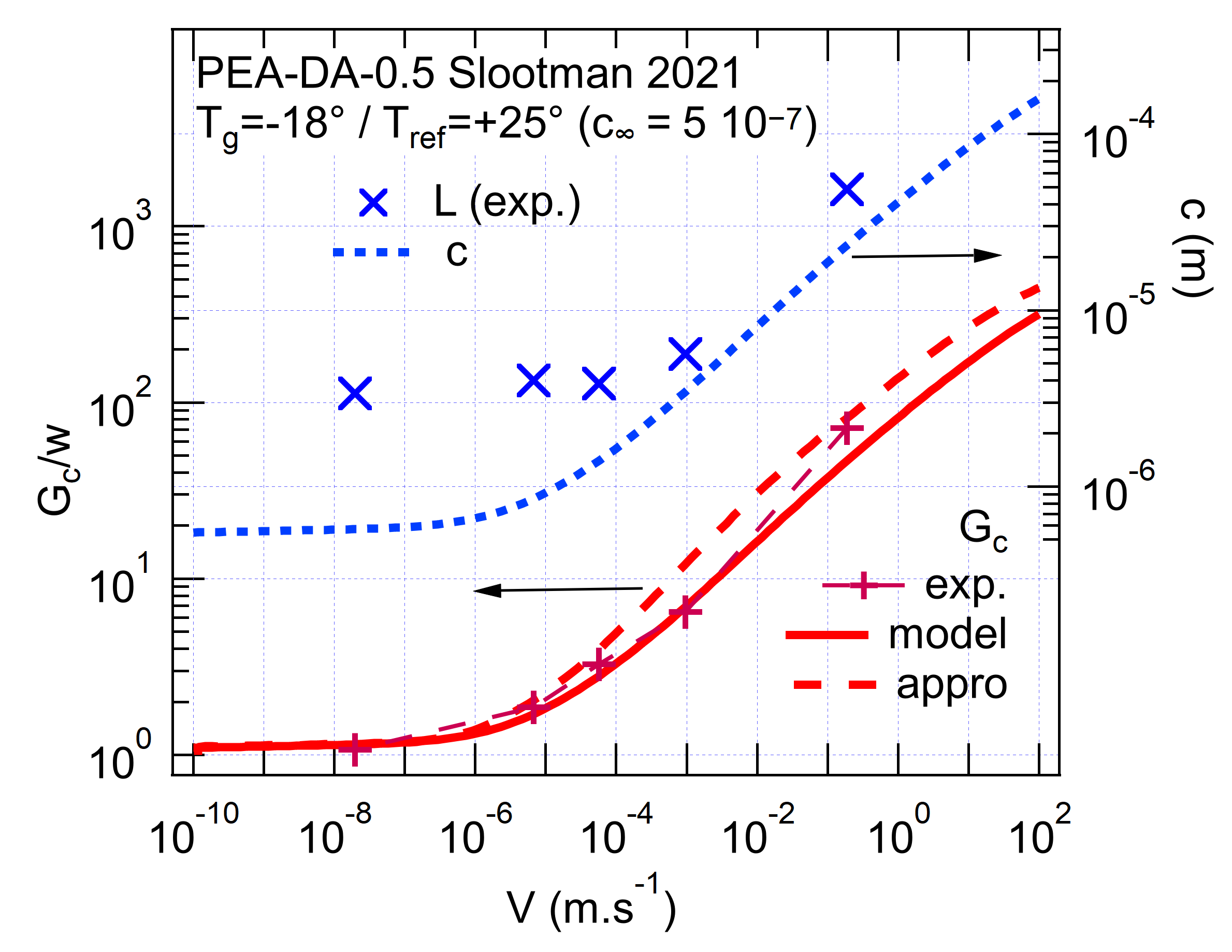}
\caption{Fracture energy of an ethylacrylate elastomer. Data taken from~\cite{Slootman2020}. a) left axis - real and imaginary parts of the shear modulus measured as a function of angular frequency - $\tan(\delta)$ is also shown (inner y axis scale). Right axis - real and imaginary parts of the shear compliance as a function of angular frequency. b) left axis - fracture energy as a function of velocity and model for $c_\infty=5\ 10^{-7}$. The result for the approximation Eq.~\ref{eq:J1_appro} is also shown as dashed lines. Right axis - size of the damage zone $L$ measured in ref~\cite{Slootman2020} (crosses) and evolution of the cohesive zone size $c$ as a function of velocity as derived from the present model (dashed line).}\label{Fig:J_G_Slootman_EA}
\end{center}
\end{figure}

As a third example, we turn to recent data documenting the rupture of a polyethylacrylate elastomer with a somewhat higher $T_g = -18^\circ$~\cite{Slootman2020}. The $G(V)$ dependence is shown in Figs.~\ref{Fig:J_G_Slootman_EA} b. Here, because $T_g$ is closer to room temperature, the determination of the low velocity adhesion was more difficult. However, the authors observe that the obtained value ($w$=30~J/m$^2$) is consistent with the Lake-Thomas prediction for this cross link density~\cite{Slootman2020}. The loss and storage modulus were also measured as a function of frequency (Figs.~\ref{Fig:J_G_Slootman_EA} a). But the most salient feature in these experiments is the use of mechanochromic moeities inserted in the polymer chains. These molecules break at a definite local stress level, an event which can be subsequently detected by fluorescence microspectroscopy. Monitoring the process \emph{post mortem}, the authors could identify the region around the crack path where material damage has occurred and measure its size $L$ (Figs.~\ref{Fig:J_G_Slootman_EA} b, crosses). It stands to reason that the size of this damage zone should somehow be related to the size of the process zone $c$ as defined here so that $L$ is an interesting addition to our rupture data.

Comparing the predicted $G(V)$ with the data for this ethylacrylate elastomer (Fig.~\ref{Fig:J_G_Slootman_EA} b, left hand axis), we find a reasonable match for $c_\infty=5\ 10^{-7}$~m. The predicted process zone size evolves with velocity between 0.6 and 30 microns (Fig.~\ref{Fig:J_G_Slootman_EA} b, dashed line, right hand axis). These values are much larger and more physically acceptable than the cases above. Moreover, the size of the damage zone $L$ as measured in the mecanofluorescence experiments is found to increase from 3 to 50 $\mu$m which is qualitatively consistent with the calculated process zone size $c$. Note that these are optical measurements and the values of $L$ are expected to level off at a minimum of ca 1 $\mu$m at low velocities due to instrumental resolution. All in all, it seems that this case is much more aptly accounted for by the linear viscoelastic theory.
% -----------------------------------------------------
\begin{figure}
\begin{center}
a) \includegraphics[width=.85\textwidth]{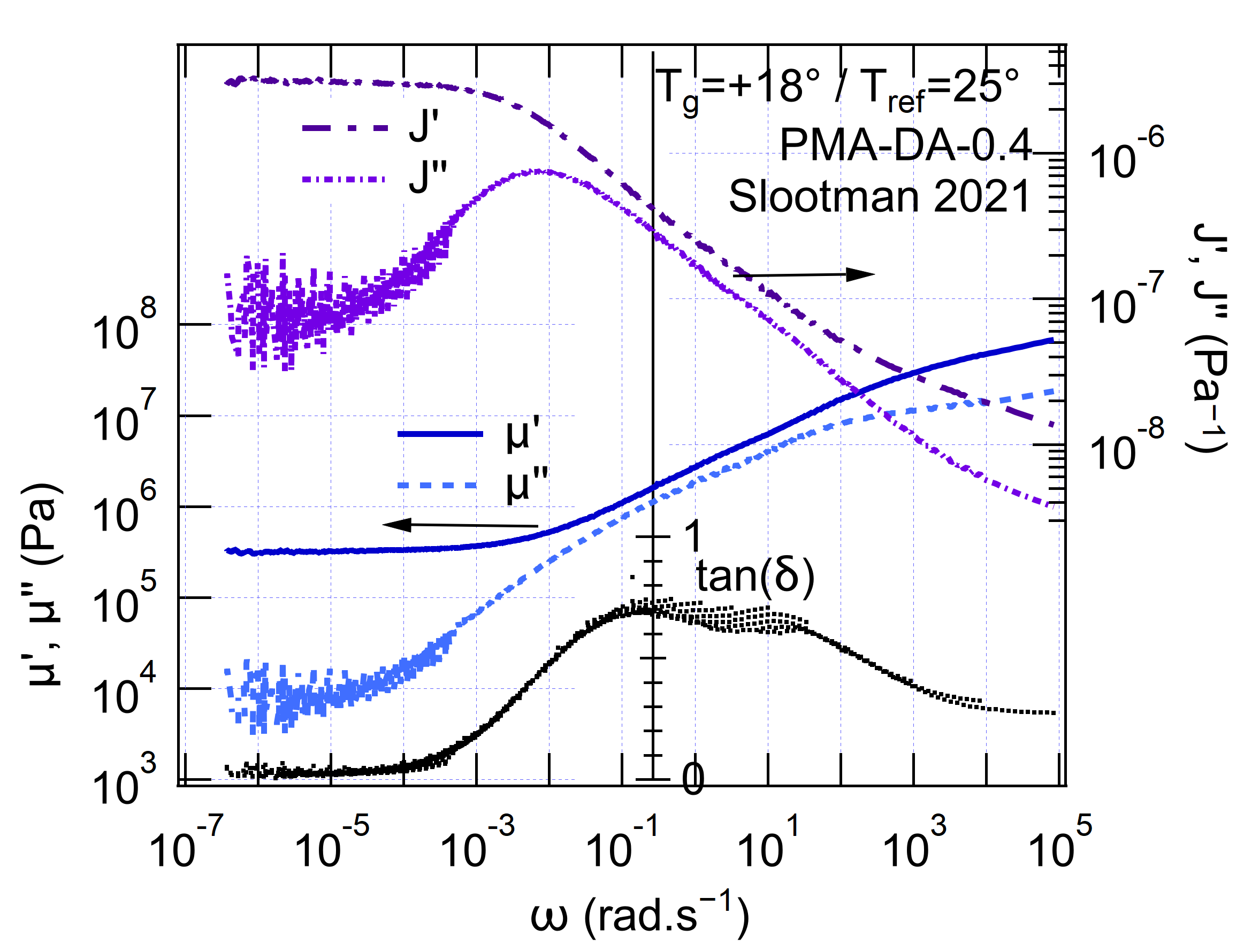}
b) \includegraphics[width=.85\textwidth]{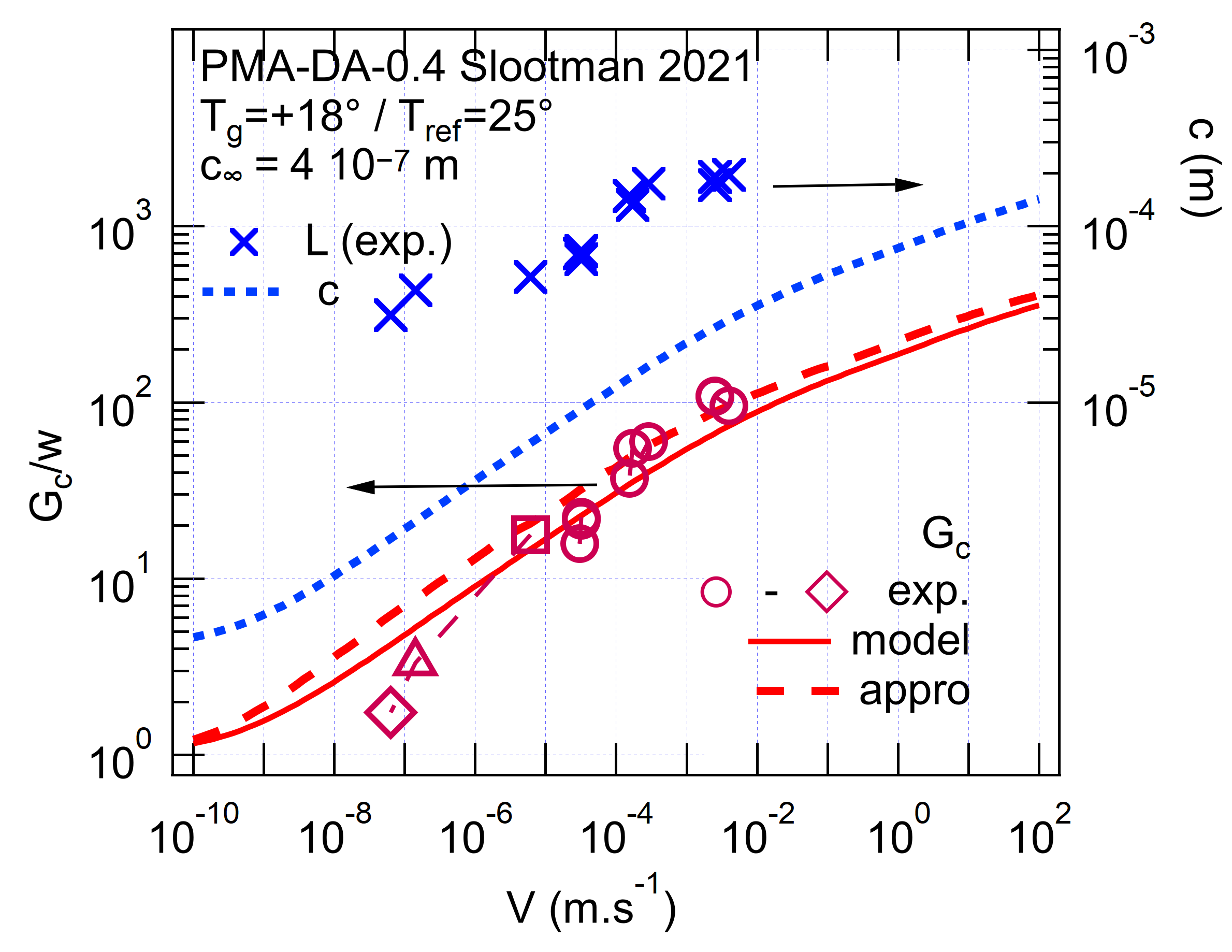}
\caption{Fracture energy of a viscoelastic methylacrylate elastomer. Data taken from~\cite{Slootman2020}. a) left axis - real and imaginary parts of the shear modulus measured as a function of angular frequency - $\tan(\delta)$ is also shown (inner y axis scale). Right axis - real and imaginary parts of the shear compliance shown as a function of angular frequency. b) left axis - fracture energy as a function of velocity ($\circ$ 25$^\circ$, $\square$ 40$^\circ$, $\triangle$ 60$^\circ$ and $\diamond$ 80$^\circ$) and model for (b) $c_\infty=4\ 10^{-7}$. The result for the approximation to the model (Eq.~\ref{eq:J1_appro}) is also shown as dashed lines. Right axis - size of the damage zone $L$ measured in ref~\cite{Slootman2020} ($\times$) and evolution of the cohesive zone size $c$ as a function of velocity as derived from the present model (dashes).}\label{Fig:J_G_Slootman_MA}
\end{center}
\end{figure}
% --------------------------------------------------------
% ----------------------------------------------------------
\begin{figure}
\begin{center}
a) \includegraphics[width=.85\textwidth]{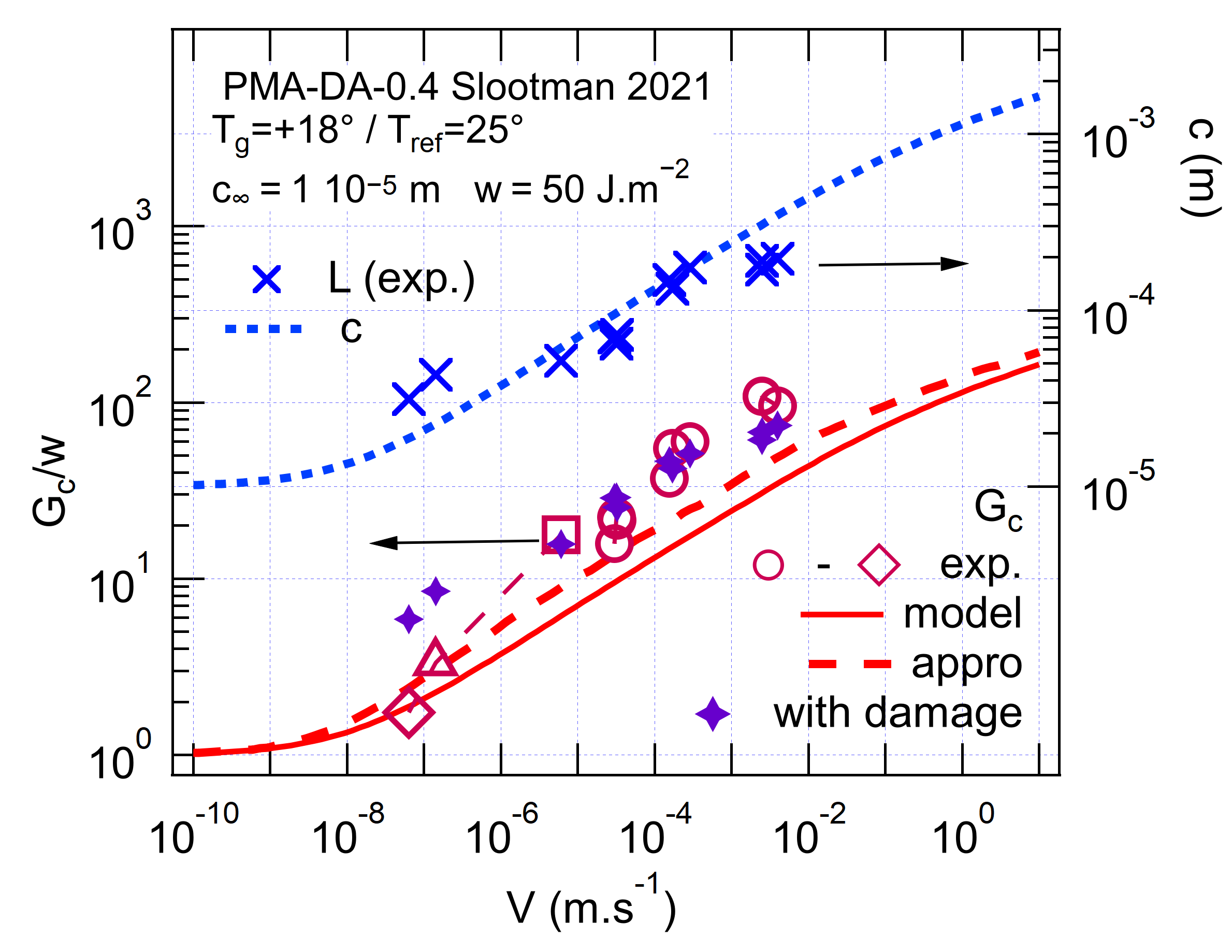}
b) \includegraphics[width=.85\textwidth]{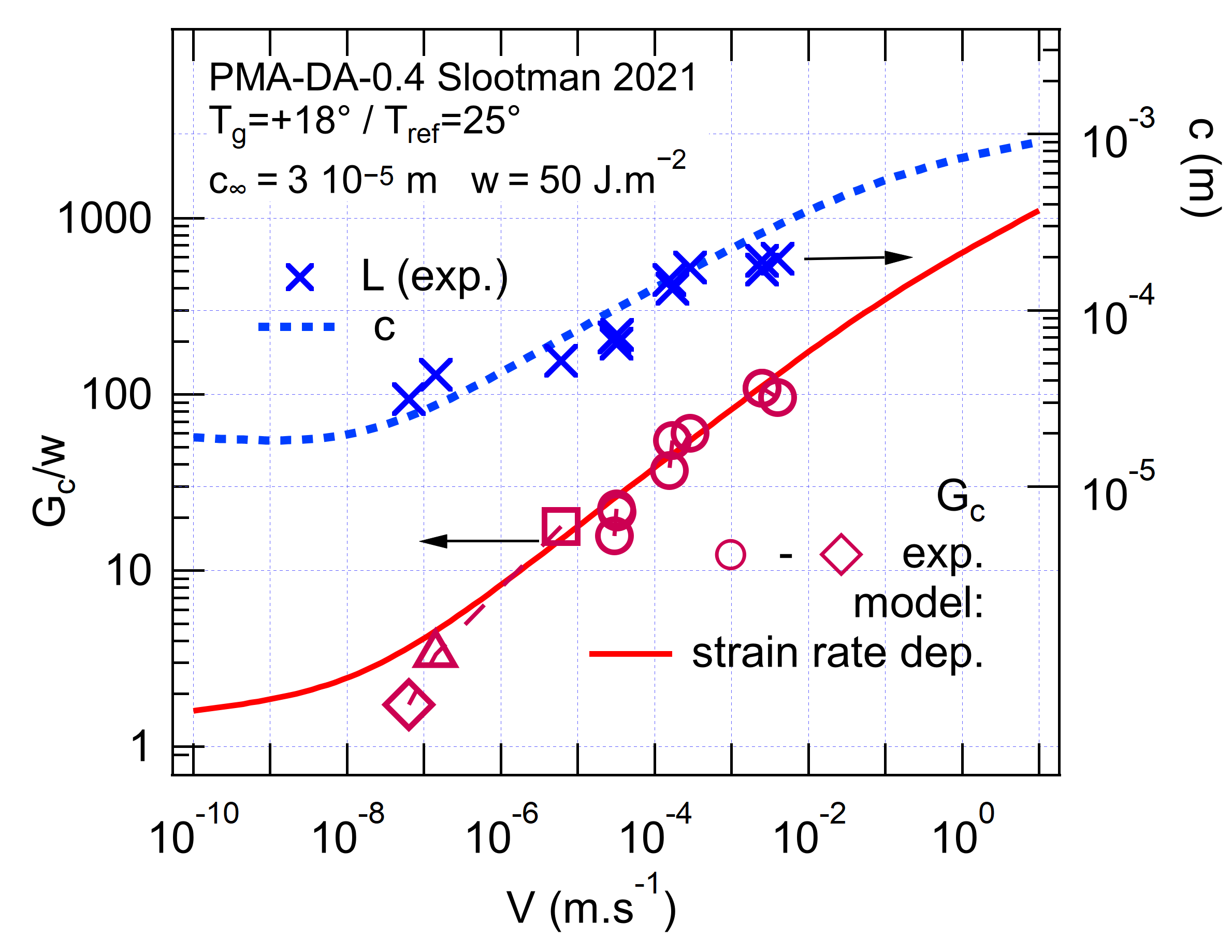}
\caption{Same data as Fig.~\ref{Fig:J_G_Slootman_MA}. a) same model as Fig.~\ref{Fig:J_G_Slootman_MA} b, with a larger process zone size $c_\infty=1\ 10^{-5}$: the linear viscoelastic contribution is smaller (plain red line). We can then add the damage energy calculated in~\cite{Slootman2020} with a rupture energy of 2 eV per bond (purple crosses). Note that the prediction for $c$ and $G$ are proportional. This is due to the classical assumption of constant cohesive zone parameters $w$ and $\sigma_0$; b) with velocity dependent $w(t_r)$ and $\sigma(t_r)$. Note that the theory now predicts non proportional evolutions of $c$ and $G$, which matches the data better.}\label{Fig:J_G_Slootman_MA_2}
\end{center}
\end{figure}
% -----------------------------------------------------------

As a last example, we consider similar experiments performed on a methylacrylate network with a glass transition temperature T$_g$=+18$^\circ$, near room temperature (\cite{Slootman2020}). For this very viscoelastic material, it is even more difficult to obtain $w$. Higher temperature results have been shifted to lower values, based on the DMA shift factors (Fig.~\ref{Fig:J_G_Slootman_MA} b). Nevertheless, there is no sign of low velocity plateau from which $w=G(V\simeq 0)$ could be evaluated. The authors suggest to take an approximate value in the same range as the ethylacrylate (see above) and we will take $w\simeq$~50 Jm$^{-2}$. The loss and storage moduli have also been measured (Fig.~\ref{Fig:J_G_Slootman_MA} a). From these data, a reasonable fit to $G(V)$ is obtained for $c_\infty=4\ 10^{-7}$~m (Fig.~\ref{Fig:J_G_Slootman_MA} b), quite similar to the value for the ethylacrylate sample. Here, two comments are in order: 1) the measured size of the damage zone $L$ (crosses) is roughly one order of magnitude larger than the size predicted for the process zone; 2) the measured power law for $G(v)$ is not accounted for properly by the theory, especially in the low velocity (\emph{i.e.} high temperature) region: the exponent found from the data is significantly larger than the predicted value. Other reasonable values of $w$ were tested and did not significantly change the predicted slope.

Since our model for the linear viscoelastic contribution is robustly based on extensive DMA data, we can relax some of the simplifying assumptions. As a first try, we note that in \cite{Slootman2020}, the authors take into account the contribution of molecular damage to the dissipated energy and hence to the effective adhesion. Assuming that the damage energy is additive and given as in \cite{Slootman2020}, we clearly need to decrease the predicted viscoelastic contribution. This can indeed be obtained  (Fig.~\ref{Fig:J_G_Slootman_MA_2} a plain red line) assuming a larger process zone size $c_\infty=1\ 10^{-5}$~m, resulting in lower strain rates for a given crack velocity. With the damage contribution inferred from \cite{Slootman2020} (damage energy 2~eV per bond), the total adhesion enhancement (purple crosses) is seen to match the data reasonably well. Moreover, the predicted evolution of the process zone size (dashed blue line) now matches the measured damage zone size much better. In this way, additive damage energy can be taken into account and in the present case seems to provide a more consistent description of the data. Agreement remains wanting in the lower velocity range, however, with some discrepancy for both process zone size and effective adhesion energy.

In fact, this shortcoming points to a classical assumption in the linear viscoelastic fracture theory: implicitly, it is assumed that the interfacial parameters $w$ and $\sigma_0$ are independant of velocity. Then, comparison of Eq.~\ref{eq:cohesive_zone_viscoelastic} with~\ref{eq:effective_adhesion_receding} shows that the predicted evolutions of $c(V)$ and $G(V)$ are strictly proportional as seen \emph{e.g.} in Fig.~\ref{Fig:J_G_Slootman_MA_2} a. Clearly the data tells a different story: the process zone size evolves more slowly than the effective adhesion energy.

To break down this strict proportionality, we no longer assume constant cohesive zone parameters $w$ and $\sigma_0$. Indeed, while strain rate independence makes sense to showcase the linear viscoelastic contribution to dissipation when comparing theories with each other, it appears simplistic when the aim is to asses data. After all, the process zone is host to large strain, high strain rate, non reversible deformations (including but not limited to non linear viscoelasticity and damage) leading to actual material separation/rupture, and these can hardly be expected to remain completely independent of strain rate. Therefore, as a second try, we now lift this restriction by allowing strain rate dependence for the process zone parameters, in the spirit of~\cite{Landis00}. Taking $w(t_r)=w\tilde w(t_r)$ and $\sigma(t_r)=\sigma_0\tilde \sigma(t_r)$, the self consistency equation Eq.~\ref{eq:cohesive_zone_viscoelastic} becomes
\begin{equation}\label{eq:evol_dep_c}
    \frac{c(t_r)}{c_\infty} = \frac{J_1(t_r=\infty)}{J_1(t_r)} \frac{\tilde w(t_r)}{\tilde\sigma(t_r)^2} 
\end{equation}
while Eq.~\ref{eq:effective_adhesion_receding} becomes
\begin{equation}\label{eq:effective_adhesion_receding_evol}
    \frac{G}{w}=\frac{J_1(t_r=\infty)}{J_1(t_r)}\tilde w(t_r) 
\end{equation}
Clearly, $c$ and $G$ are no longer proportional. To demonstrate the potential of this new point of view, we take very simple \emph{ad hoc} dependencies $\tilde \sigma = 1+(t_\alpha/t)^\alpha$ and $\tilde w = 1+(t_\beta/t)^\beta$. We observe a significant improvement in the match between model and data for $c_\infty = 3\ 10^{-5}$~m, $w=50$~Jm$^{-2}$ and weak strain rate dependences $\alpha=\beta=0.1$ and $t_\alpha=t_\beta=1\ 10^3$ (Fig.~\ref{Fig:J_G_Slootman_MA_2} b).

% ------------------------------
\begin{table}[]
\centering
\begin{tabular}{|c|c|c|c|c|}
\hline
    Exp & $\omega_{\rm{onset}}$ (s$^{-1}$) & $V_{\rm{onset}}$ (ms$^{-1}$) & $c_{\rm{appro}}$ (m) & $c_{\infty}$ (m) \\
\hline
    Tay & $3\ 10^{5}$ & $6\ 10^{-7}$ & $2\ 10^{-12}$ & $3\ 10^{-13}$ \\
    Gent SB &  $1\ 10^{-5}$ & $2\ 10^{-16}$ & $2\ 10^{-11}$ & $2\ 10^{-11}$ \\
    Slootman EA &  $4\ 10^{1}$ & $1\ 10^{-5}$ & $2.5\ 10^{-7}$ & $5\ 10^{-7}$ \\
    Slootman MA & $1\ 10^{-2}$ & $1\ 10^{-7}$ & $1\ 10^{-5}$ & $4\ 10^{-7}$ \\
\hline
\end{tabular}
\caption{Simple evaluation of the approximate process zone size $c_{\rm{appro}}=V/\omega$ where $V$ and $\omega$ are defined by $G(V)/w=J'(0)/J'(\omega)=2$) and comparison with process zone size calculated from the present model.}
\label{tab:c_appro}
\end{table}
% --------------------------------
    
\section{Discussion}\label{sec:discussion}
The framework given by Eqs.~\ref{eq:definition_V} (crack velocity), \ref{eq:cohesive_zone_viscoelastic} (self-consistency) and~\ref{eq:effective_adhesion_receding} (relation between far field and process zone) is not more than the umpteenth variation over the same linear viscoelastic model~\cite{Knauss15_reviewfractureviscoelastic,Hui22}. However, with the effective crack compliance cast under the form Eq.~\ref{eq:J1_freq}, it allows a very direct comparison with data. A number of comments are in order, based on the examples we have just analysed.
\subsection{Quality of the data}
The comparison with the data is reliable if the measured $G(V)$ includes not only the intermediate velocity dependent regime but also some evaluation of $G$ in the low velocity regime, as in Sec.~\ref{sec:Tay_data}. From a low velocity plateau we can derive the interfacial rupture energy $w$ and from the fit to the velocity dependent regime we can determine the remaining free parameter $\sigma_0$ (or equivalently $c_\infty$). If the low velovity plateau is not evidenced in the data, as is the often case, $w$ is not determined and the analysis is essentially inconclusive. Indeed, $w$ and $\sigma_0$ are then equivalent. Figs.~\ref{Fig:J_G_Tay} and \ref{Fig:J_G_Gent}, which show G(V) calculated for various $c_\infty$ values, illustrate this problem very clearly: in the intermediate region, where $G$ depends upon velocity, changing $c_\infty$ mainly results in a vertical shift of the curve -- this is also what happens when changing $w$. In brief, without a proper evaluation of $w$, \emph{e.g.} from a low velocity plateau, a reliable evaluation of the linear viscoelastic contribution becomes impossible and no conclusion can be derived as to the applicability of the model. Direct comparison with $\mu''(\omega)$ through power law exponents, for example, does not bring any insight: different exponents can be found in different frequency ranges (Fig.~\ref{fig:Tay_GpGpp}) and a comparison with $\mu''$ has actually little physical ground. 

\subsection{Comparison with Gent's approach}
In depth data analysis has also been carried out by Gent in his 1994 evaluation of elastomer rupture~\cite{Gent94} using a more qualitative approach. We now compare our approach with Gent's.

Gent first comments on the evolution of $G(V)$, which resembles $\mu'(\omega)$ but not $\mu''(\omega)$ nor $\tan(\delta)$ ("only the storage modulus varies with frequency in approximately the right way and approximately the right factor"). He then proceeds to use a simplified approach based on the classical Schapery/Knauss formulation~\cite{Schapery75a,Bowen1992a}. To that aim, he uses the relation
\begin{equation}\label{eq:effective_adhesion_receding_Gent}
\frac{G(V)}{w}=\frac{\mu'(t_r=c/V)}{\mu'(t_r=\infty)} 
\end{equation} 
Comparing to Eq.~\ref{eq:effective_adhesion_receding}, we find that here $1/\mu'(t_r)$ stands for a simple approximation to the effective compliance $J_1(t_r)$.

This comparison suggests a further approximation: replacing Eq.~\ref{eq:J1_freq} with
\begin{equation}\label{eq:J1_appro}
    J_1(t_r)\simeq J'(\omega=2\pi/t_r)
\end{equation}
We have also plotted the results (dashed lines) in Figs.~\ref{Fig:J_G_Tay}, \ref{Fig:J_G_Gent}, \ref{Fig:J_G_Slootman_EA} and \ref{Fig:J_G_Slootman_MA}. It appears that this very simple approximation is working well, but the increase of $G(V)$ with velocity is anticipated by roughly one decade in velocity. This can be understood if we recall that, due to the convolution of $J'(\omega)$ in Eq.~\ref{eq:J1_freq},  $J_1(\omega)$ follows $J'(\omega)$ with a frequency lag of the order of a decade (Fig.~\ref{fig:Jp_J1}). Note that such a shift of the characteristic time, for the same level of approximation, has been observed earlier, \emph{e.g.} in the time domain model by Greenwood, who found a scaling of 0.24 between the two characteristic times, assuming a specific crack face interaction and a three-element solid~\cite{Greenwood04}.

In brief, the approximation Eq.~\ref{eq:J1_appro} is certainly quite acceptable as a first approach, for example when checking if the process zone is ludicrously small... Note also that $J'(\omega)$ can be reasonably well approximated by $1/\mu'(\omega)$, especially at frequencies below the glass transition where $\mu''(\omega)\ll \mu'(\omega)$ so that there is a close connection between Eq.~\ref{eq:effective_adhesion_receding_Gent} on the one hand, and Eqs.~\ref{eq:effective_adhesion_receding} and \ref{eq:J1_appro} on the other hand. Applying the procedure outlined by Gent~\cite{Gent94}, but using Eq.~\ref{eq:J1_appro}, we can now readily derive an approximate process zone size $c_{\rm{appro}}$ and predict an effective rupture/adhesion energy from the data, without performing the convolution in Eq.~\ref{eq:J1_freq}. Consider Eq.~\ref{eq:effective_adhesion_receding} and take a given value for the ratio, say an amplification of 2. Define $V_{\rm{onset}}$ and $\omega_{\rm{onset}}$ by $G(V_{\rm{onset}})/w=J'(\omega=0)/J'(\omega_{\rm{onset}}) = 2$. Basically, we are here taking the velocity at the upturn of $G(V)$, the angular frequency at the downturn of $J'(\omega)$ and evaluating $c_{\rm{appro}}$ as $V_{\rm{onset}}/\omega_{\rm{onset}}$. Results from this procedure are shown in Table~\ref{tab:c_appro} : this simple approach efficiently predicts $c_\infty$ in order of magnitude and, in favourable cases (Gent SB, Slootman EA) even reasonably well. In particular, the much larger values predicted for the acrylate elastomers readily come out, in fair agreement with the observed extent of damage $L$~\cite{Slootman2020}. We conclude that, given its simplicity, Eq.~\ref{eq:J1_appro} is definitely a good way to obtain a quick evaluation of the contribution of linear viscoelasticity to rupture/adhesion. With differences of up to a factor of ten between the two evalutions of $c_\infty$ (cf the Tay and the Slootman MA data) however, using the full model may still be a good idea for finer analyses, or for more elaborate cases.

\subsection{Evaluation of the cohesive stress}
It is also interesting to evaluate the cohesive stress $\sigma_0$ from the $c$ values. The results are summarized in Table~\ref{tab:stress}. The anomalously small cohesive zone sizes result in cohesive stresses which exceed the measured Young's modulus by two to three orders of magnitude. In contrast, the cohesive stress for the acrylates is "only" between 3 to 20 times larger than the small strain, relaxed Young's modulus. These values are much more acceptable, all the more so as the cohesive stress reflects the non-linear phenomena taking place in the process zone, with large strains and strain rates, leading to rupture.

\subsection{When does linear viscoelasticity play a role?}
Finally, because it is simple and robust, the model propounded here can be used to pertinently assess the contribution of linear viscoelasticity to the fracture/adhesion energy of soft solids in a variety of cases. Based on the four examples shown above, it is clear that standard elastomers are not good candidates: this is the conclusion reached by Gent and others~\cite{Gent94,Knauss15_reviewfractureviscoelastic,Hui22}. In this case, the glass transition is at such high frequency (Fig~\ref{fig:Jp_J1}) that implausibly small process zone sizes must be invoked to account for the measured fracture/adhesion energies at typical crack velocities. This means that in elastomers, the contribution of linear viscoelasticity remains minute and dissipation is dominated by non-linear processes connected to damage and rupture, in a process zone with a size too large to induce significant linear viscoelastic dissipation.

In contrast, we have also considered acrylates with glass transitions closer to room temperature: significant viscoelastic dissipation is predicted for process zone sizes in the range of microns to tens of microns, consistent with the damage zone sizes measured by mecanoluminescence. The measured velocity dependence of the fracture energy can then be accounted for within the scope of a linear model, with physically acceptable estimates of the cohesive stresses and process zone sizes.
% ---------------------------------------------------
\begin{table}[]
    \centering
    \begin{tabular}{|c|c|c|c|c|c|}
    \hline
        Exp & $w$ (J.m$^{-2}$) & $T_g$ ($^\circ$ C) & $c_\infty$ (m) & $J(\infty)$ (Pa$^{-1}$) & $\sigma_0 $ (Pa) \\
    \hline
        Tay &  0.043 & -72 & $3\ 10^{-13}$ & $3.1\ 10^{-6}$ &  $3.0\ 10^{8}$ \\
        Gent SB &  2.3 & -27 & $2\ 10^{-11}$ & $1.7\ 10^{-6}$ &  $3.7\ 10^{8}$ \\
        Slootman EA &  30 &-18 &  $5\ 10^{-7}$ & $3.0\ 10^{-6}$ &  $6.3\ 10^{6}$ \\
        Slootman MA &   50 & +18 & $4\ 10^{-7}$ & $3.1\ 10^{-6}$ &  $9.0\ 10^{6}$ \\
        Sl. MA (with damage) & " & " & $1\ 10^{-5}$ & " &  $1.8\ 10^{6}$ \\
        Sl. MA (var. $w$ and $\sigma$) & " & " & $3\ 10^{-5}$ & " &  $1.0\ 10^{6}$ \\
    \hline
    \end{tabular}
    \caption{Model parameters and characteristic stress $\sigma_0$ for different experiments as calculated from Eq.~\ref{eq:cohesive_zone_viscoelastic}.}
    \label{tab:stress}
\end{table}
% ----------------------------------------------------------

An interesting feature in these data sets is the size of the damage zone surrounding the crack path measured by mecanoluminescence. Comparison of the predictions with the data directly challenges the ubiquitous (and highly suspect) assumption that the process zone parameters are independent of velocity. Of course, the interpretation of the measured damage zone size is not unambiguous and the precise relation that exists between this experimental damage zone and the process zone in the model is not clear. Here we have assumed that they are equal, and we could show that need only a weak dependence of both interfacial rupture energy and cohesive stress to account for the data. This result could not be obtained in the standard framework where the process zone parameters are constant. 

As a final observation, we consider the role of non-linear processes in the rupture/adhesion of soft solids. Note that a time-temperature rescaling is often used to patch together rupture/adhesion data taken at different temperatures to build up a master curve over an extensive velocity range, as was done \emph{e.g.} in Refs.~\cite{Gent94} or \cite{Slootman2020}. For a linear theory, the reference temperature is indifferent, and the results should not be affected. But the rupture phenomena are inherently non linear so that the use of time temperature superposition raises questions. Maybe the present results mean that scaling too far away from the measurement temperature leads to difficulties. We suggest that this is the reason why the linear viscoelastic crack theory could work for... viscoelastic systems, but seems inadequate for \emph{bona fide} elastomers.

\section{Conclusion}
We have analysed rupture/adhesion data for more or less viscoelastic elastomers. To do so, we have first provided a model relating the velocity dependence of the effective rupture/adhesion energy and the linear mechanical response (loss and storage modulus) as a function of frequency. In this model, an effective crack compliance $J_1$ encapsulates a self-consistent description of the viscoelastic deformation taking place around the process zone (Eq.~\ref{eq:cohesive_zone_viscoelastic}). It can be connected to the far field to provide an expression for the linear viscoelastic contribution to the effective rupture/adhesion energy (Eq.~\ref{eq:effective_adhesion_receding}).

$J_1$ can be expressed in a form (Eq.~\ref{eq:J1_freq}) especially suitable for data analysis (Fig.~\ref{fig:Jp_J1}). Alternatively, a very simple approximation for $J_1$, Eq.~\ref{eq:J1_appro}, has also been given and was shown to work reasonably well, although it is shifted in frequency by about one decade. This simple approximation also provides a direct connection to Gent's approach~\cite{Gent94}. 

As already pointed out previously~\cite{Gent96,Knauss15_reviewfractureviscoelastic,Hui22}, for typical elastomers, the data can be adjusted only by assuming ridiculously small values of $c_{\infty}$ (and ridiculously high values of the cohesive stress), which actually means that the linear viscoelastic dissipation is negligible in this case. However, for the two more viscoelastic systems investigated here, the values of $c_{\infty}$ are much more acceptable, and consistent with the damage zone sizes measured by mecanoluminescence. Moreover, for the methacrylate system, for which the data are more complete, we find that we can account for the full set of data only if the simplistic assumption of velocity independent process zone parameters is abandoned. A simple (weak) phenomenological dependence for cohesive energy and cohesive stress on strain rate was shown to provide significantly improved agreement with the data.

We have also highlighted that good quality data should contain not only the velocity dependent regime but also the low velocity plateau from which the interfacial rupture energy $w$ can be evaluated. In this case, the model unequivocally leaves one single adjustable parameter, directly related to the cohesive stress. In the absence of reasonable estimate for $w$, no conclusive analysis can be carried out.

In brief, the present description of linear viscoelastic fracture is simple and powerful enough that it can help understand the fracture of soft solids. By providing a simple prediction for the linear viscoelastic contribution to the rupture/adhesion energy under a form which can be directly applied to data, it should help reach a better understanding of rupture in soft matter.
\section*{Acknowledgements}
For several insightful suggestions and in particular for pointing me towards Eq.~\ref{Eq_Jp_freq2} and the high quality data in ref~\cite{Tay06}, I warmly thank my colleague Christian Frétigny, who characteristically declined my invitation to co-author the paper. I am very grateful to the late H. Montes who graciously provided the data for figure~\ref{fig:Tay_GpGpp} which complements ref~\cite{Tay06}. I am also indebted to Matteo Ciccotti and C. Y. (Herbert) Hui for interesting discussions over this topic and to Costantino Creton and Jean Comtet for their comments and for sharing the data from ref.~\cite{Slootman2020}.

The data used for the calculations and a Matlab code are available as supplementary materials.
\bibliographystyle{unsrt}
\bibliography{fracture,Contact_and_adhesion}
\appendix
\section{Approximate derivation of the self-consistency equation}\label{sec:appendix_self_con}
For completeness, we sketch a derivation of Eq.~\ref{eq:Cohesive_Zone_Size_Elastic_}. which expresses the necessary self-consistency of the stress distribution in the cohesive zone.
This relation directly results from the integration of the cohesive stresses $\sigma$ in the displacement of the opening crack faces $u_z$ \emph{ie} $$w=-\int_{0}^{+\infty}\sigma(z) du_z$$
which can also be written as
\begin{equation}\label{Eq_Self_consistent_generic}
w=-\int_{0}^{c}\sigma\frac{du_z}{dx}dx
\end{equation}
A simple scaling relation is obtained by noting that
\begin{equation}\label{Eq_Cohesive_zone_material_response}
du_z/dx\simeq \sigma_0/E^*\
\ \ \ \ \hbox{(elastic)}
\end{equation}
from which an approximate form of Eq.~\ref{eq:Cohesive_Zone_Size_Elastic_} is derived directly. The exact prefactor, of order one depends on the details of the cohesive model used. 

\section{Effective compliance in the frequency domain}\label{sec:appendix_frequency_domain}
In the following, we will use a different approximation for the effective crack tip compliance. In our previous work~\cite{Barthel09,Haiat07}, we found the approximate expression
\begin{equation}\label{eq:Effective_Compliance_Opening}
{J}_{1}(t_r)=\frac{2}{{t_r}^2}\int_0^{t_r}(t_r-\tau)J(\tau)d\tau\ \ \ \ \ %\hbox{(opening crack)}
\end{equation}
for the effective compliance of the crack.
Eq.~\ref{eq:Effective_Compliance_Opening} is not practical for fast materials like natural rubber. Indeed, experimentally, good small strain dynamic data is obtained with time temperature superposition in the frequency domain. It takes the form of a complex shear modulus $\mu(\omega) = \mu'(\omega)+i\mu''(\omega)$ where $\omega=2\pi\nu$ is the angular frequency. In Fourier space, this complex shear modulus is the inverse of the complex compliance $J(\omega) = J'(\omega)+i J''(\omega)$
\begin{equation}\label{Eq_J_mu}
    J(\omega)= \mu(\omega)^{-1}
\end{equation}
while $J(\omega)$ is the Fourier transform
\begin{equation}\label{Eq_Jp_freq2}
J(t) = \frac{2}{\pi}\int_0^{+\infty} J'(\omega)\sin(\omega t)d\left(ln(\omega)\right)
\end{equation}

Inserting this equation into the time domain expression for the local effective compliance Eq.~\ref{eq:Effective_Compliance_Opening}, we obtain the frequency domain expression Eq.~\ref{eq:J1_freq}.

\end{document}